\documentclass[10pt]{article}
\usepackage{amsmath,amssymb}
\usepackage{amsthm}
\usepackage{mathptmx}
\usepackage{cite}
\usepackage{latexsym}

\numberwithin{equation}{section}

\usepackage{hyperref}

\newtheorem{theorem}{Theorem} 

\newtheorem{proposition}{Proposition}

\begin{document}

\title{Bounds on M/R for Charged Objects with positive Cosmological constant}
\author{H{\aa}kan Andr\'{e}asson\\
Mathematical Sciences\\ University of Gothenburg\\
        Chalmers University of Technology\\
        S-41296 G\"oteborg, Sweden\\
        email: hand@chalmers.se\\
        \ \\
        Christian G.~B\"ohmer\\
        Department of Mathematics and Institute of Origins\\ University College London\\
        Gower Street, London, WC1E 6BT, UK\\
        email: c.boehmer@ucl.ac.uk\\
        \ \\
        Atifah Mussa\\
        Department of Mathematics and Institute of Origins\\ University College London\\
        Gower Street, London, WC1E 6BT, UK\\
        email: atifahm@math.ucl.ac.uk}

\date{\today}
\maketitle

\begin{abstract}
We consider charged spherically symmetric static solutions of the Einstein-Maxwell equations with a positive cosmological constant $\Lambda.$
If $r$ denotes the area radius, $m_g$ and $q$ the gravitational mass and charge of a sphere with area radius $r$ respectively, we find that for any solution which satisfies the condition $p+2p_{\perp}\leq\rho,$ where $p\geq 0$ and $p_{\perp}$ are the radial and tangential pressures respectively, $\rho\geq 0$ is the energy density, and for which $0\leq \frac{q^2}{r^2}+\Lambda r^2\leq 1,$ the inequality

\begin{equation*}
\frac{m_g}{r} \leq \frac{2}{9}+\frac{q^2}{3r^2}-\frac{\Lambda r^2}{3}+\frac{2}{9}\sqrt{1+\frac{3q^2}{r^2}+3\Lambda r^2}
\end{equation*}
holds. We also investigate the issue of sharpness, and we show that the inequality is sharp in a few cases but generally this question is open.
\end{abstract}

\section{Introduction}
An important question is to determine an upper bound on the gravitational red shift of spherically symmetric static objects. In the case with vanishing cosmological constant and charge this is equivalent to determining an upper bound on the compactness ratio $M/R,$ where $M$ is the ADM mass and $R$ the area radius of the boundary of the static object. Buchdahl's theorem~\cite{Bu} is well-known and shows that a spherically symmetric isotropic object for which the energy density is non-increasing outwards satisfies the bound
\begin{align}\label{Buchdahl}
\frac{M}{R}\leq\frac{4}{9}.
\end{align}
The inequality is sharp, but the solution which saturates the inequality within the class of solutions considered by Buchdahl violates the dominant energy condition and is therefore unphysical. Moreover, the assumptions that the pressure is isotropic, and the energy density is non-increasing, are quite restrictive. In \cite{An1} it was shown that the bound (\ref{Buchdahl}) holds generally, i.e., independently of the Buchdahl assumptions, for the class of solutions which satisfy the condition
\begin{align}\label{energycondition}
p+2p_{\perp}\leq \rho,
\end{align}
where $p\geq 0$ is the radial pressure, $p_{\perp}$ the tangential pressure and $\rho\geq 0$ the energy density. This condition implies in particular that the dominant energy condition holds. In addition it was shown that the inequality is sharp and that the saturating solution is unique. An alternative proof was given in \cite{KS} where more general conditions than (\ref{energycondition}) were treated but the uniqueness of the saturating solution was not settled. The inequality derived in~\cite{An1} also holds inside the object and the inequality then takes the form $m(r)/r\leq 4/9,$ where $m=m(r)$ is the mass within the sphere of area radius $r$. If charged spheres are considered the corresponding inequality also involves the charge $q=q(r)$ and it was shown in~\cite{An5} that the inequality generalizes to
\begin{align}\label{HA}
\frac{\sqrt{m_g}}{\sqrt{r}}\leq\frac{1}{3}+\sqrt{\frac{1}{9}+\frac{q^2}{3r^2}},
\end{align}
where $m_g=m_g(r)$ is the gravitational mass, cf.~\cite{An5}.
The inequality (\ref{HA}) is sharp but in~\cite{AER} numerical evidence is given, in the case of the Einstein-Vlasov-Maxwell system, that the saturating solution is non-unique.

In the case without charge, the inclusion of a positive cosmological constant $\Lambda$ was investigated in~\cite{AB}. If $0\leq \Lambda r^2\leq 1,$ the following inequality was obtained
\begin{align}\label{AB}
\frac{m}{r}\leq\frac{2}{9}-\frac{\Lambda r^2}{3}+\frac{2}{9}\sqrt{1+3\Lambda r^2}.
\end{align}
In this case the question of sharpness was not settled except in the degenerate cases when $\Lambda r=0$ or $\Lambda r^2=1$, cf. Remark 1 below for an interpretation of the former case.

Bounds on mass-radius ratios can also be obtained by considering special solutions or following Buchdahl's original approach. These derivations can neither settle sharpness nor uniqueness, however, the results obtained are often surprisingly similar, see~\cite{Boehmer:2003uz,Boehmer:2006ye,Boehmer:2007gq,BM,MDH}.

In the present study we include both charge and a positive cosmological constant. Under the condition that $0\leq \frac{q^2}{r^2}+\Lambda r^2 \leq 1,$ we derive the following inequality
\begin{equation*}
\frac{m_g}{r}\leq\frac{2}{9}+\frac{q^2}{3r^2}-\frac{\Lambda r^2}{3}+\frac{2}{9}\sqrt{1+\frac{3q^2}{r^2}+3\Lambda r^2}.
\end{equation*}
We also address the question of sharpness by considering infinitely thin shell solutions since these saturate the inequality in the absence of a cosmological constant. We show that when $\Lambda>0$ infinitely thin shells do not saturate the inequality except in the degenerate cases $\Lambda r=0$ and $\frac{q^2}{r^2}+\Lambda r^2=1.$ Throughout the paper we use three different mass conventions, $m=m(r)$ the mass inside a sphere of radius $r$~(\ref{m}), $M$ the ADM mass~(\ref{ADM}) and $m_g$ the gravitational mass~(\ref{mg}). Likewise, $q=q(r)$ denotes the charge inside a sphere of radius $r$ and $Q$ denote the total charge.

The outline of the paper is as follows. In the next section the system of equations is presented and our main results are stated in detail. Sections 3 and 4 are devoted to their proofs.

\section{Set up and main results}

The pressure is allowed to be anisotropic so that the radial and tangential pressures $p$ and $p_\perp$ are not necessarily equal, but we require that the following inequality holds
\begin{align}
p+2p_\perp \leq \rho\,. \label{strong}
\end{align}
Here the energy density $\rho$ and radial pressure $p$ are non-negative.
We remark that this condition always holds in the case of collisionless matter, i.e.,
for the Einstein-Vlasov system.

We will examine a charged spherically symmetric mass distribution with a non zero cosmological constant. We write the metric as
\begin{align}
  ds^2 = -e^{a(r)} dt^2 + e^{b(r)} dr^2 + r^2(d\theta^2 + \sin^2\theta d\phi^2).
  \label{metric}
\end{align}
If we take
\begin{equation*}
  e^{a(r)} = 1 - \frac{2M}{r} + \frac{Q^2}{r^2} -\frac{\Lambda r^2}{3} = e^{-b(r)},
\end{equation*}
this is the Reissner-Nordstrom-de Sitter solution, and as $r \to \infty$ this solution tends to de Sitter space.
This describes all Reissner-Nordstrom-de Sitter solutions except for a certain class of solutions called the charged Nariai solutions which are not asymptotically de Sitter, see~\cite{BM}.

In order to write down the Einstein-Maxwell equations we introduce the charge
$q=q(r)$ within a sphere of area radius $r$ given by
\begin{equation*}
  q(r) = 4\pi \int_0^r{e^{(a+b)(\eta)}\eta^2 j^0 d\eta},
\end{equation*}
where $j^0=j^0(r)$ is the charge density. The total charge is denoted by $Q$
so that $Q=q(R)$, where $r=R$ is area radius of the boundary of the object.
Given the metric~(\ref{metric}), the Einstein-Maxwell field equations take
the form
\begin{align}
  8\pi \rho + \frac{q^2}{r^4}&=\frac{1}{r^2} \frac{d}{dr}\left(r-re^{-b} \right) - \Lambda,
  \label{ee1}\\
  8\pi p - \frac{q^2}{r^4}&=\frac{e^{-b}}{r^2} + \frac{a' e^{-b}}{r} - \frac{1}{r^2}+ \Lambda,
  \label{ee2}\\
  8\pi p_\perp -\frac{q^2}{r^4}&= \frac{e^{-b}}{2}\left( a'' +\left( \frac{a'}{2}+\frac{1}{r}\right) \left(a'-b' \right)\right)+\Lambda,
  \label{ee3}\\
  F_{rt} &= \frac{e^{(a+b)/2}}{r^2}q, 
\end{align}
where $F_{rt}$ is the only non-vanishing component of the electromagnetic tensor $F_{ij}$. For more information on the derivation of these equations, in the case $\Lambda=0$, we refer to~\cite{AER}.
From these equations we obtain
the Tolman-Oppenheimer-Volkoff (TOV) equation for the pressure
\begin{equation*}
  p' + \frac{a'}{2}(\rho+p)+\frac{2}{r}(p-p_{\perp})- \frac{qq'}{4\pi r^4} = 0.
\end{equation*}
Solving (\ref{ee1}) yields
\begin{equation}\label{match}
e^{-b(r)}=1-\frac{2m(r)}{r}-\frac{f(r)}{r}-\frac{\Lambda r^2}{3},
\end{equation}
where
\begin{equation}
m(r)=\int_0^r 4\pi\,\rho\eta^2\,d\eta,
\label{m}
\end{equation}
and
\begin{equation*}
  f(r)=\int_0^r{\frac{q^2(\eta)}{\eta^2}d\eta}.
\end{equation*}
Requiring that (\ref{match}) matches the Reissner-Nordstrom-de Sitter solution
at the boundary $r=R$ of the charged object gives
\[
1-\frac{2M}{R} + \frac{Q^2}{R^2}-\frac{\Lambda R^2}{3}=1-\frac{2m(R)}{R}
-\frac{f(R)}{R}-\frac{\Lambda R^2}{3},
\]
or
\begin{equation}
M=m(R)+\frac{Q^2}{2R}+\frac{f(R)}{2},
\label{ADM}
\end{equation}
which defines the gravitational mass at $r=R$, the mass measured by a satellite
in orbit around the object. In view of this relation we define
the gravitational mass $m_g=m_g(r)$ of a sphere with area radius $r$ by
\begin{equation}
m_g(r)=m(r)+\frac{q^2(r)}{2r}+\frac{f(r)}{2}.
\label{mg}
\end{equation}
We can now formulate our main result.
\begin{theorem}
\label{thm1}
Let $\Lambda\geq 0$ and assume that a solution of the Einstein-Maxwell equations
(\ref{ee1})-(\ref{ee3}) exists which satisfies (\ref{strong}). If
\begin{equation}\label{condthm}
\frac{q(r)^2}{r^2}+\Lambda r^2\leq 1,
\end{equation}
then
\begin{equation}\label{mainineq}
  \frac{m_g}{r} \leq \frac{2}{9} + \frac{q^2}{3r^2} - \frac{\Lambda r^2}{3} + \frac{2}{9}\sqrt{1 + \frac{3q^2}{r^2} + 3\Lambda r^2}.
\end{equation}
\end{theorem}
We note that setting $\Lambda$ or $q$ to zero will result in the inequalities~(\ref{HA}) and~(\ref{AB}) respectively. It should be emphasised that in the context of static and spherically symmetric solutions, the inequality (\ref{mainineq}) is very general in the sense that there are no other reasonable modifications which can be added. It contains charge and the cosmological term and also allows for anisotropic matter. On the other hand there is room for improvements since the inequality is only shown to hold under the assumptions that $\Lambda\geq 0$ and that condition (\ref{condthm}) holds, and moreover, it is not known if it is sharp in general. This issue will now be discussed in more detail.

As was mentioned in the introduction, in the case when $Q=\Lambda=0,$ infinitely thin shell solutions saturate the inequality uniquely, cf.~\cite{An1}. In the case $Q\ne 0$ and $\Lambda=0,$ infinitely thin shell solutions also saturate the inequality as is shown in~\cite{An5}. However, in~\cite{AER} numerical evidence is given that there is also another type of saturating solution when the inner and outer horizon of the Reissner-Nordstr\"{o}m black hole coincide, and the saturating solution is thus not unique. In the case $Q=0$ and $\Lambda>0$ the issue of sharpness is investigated in~\cite{AB} and it is shown that infinitely thin shell solutions do not satisfy the inequality except in the cases $\Lambda r^2=0$ or $\Lambda r^2=1.$ In the latter situation there is also a constant energy density solution where the exterior spacetime is the Nariai solution which satisfies the inequality, hence the saturating solution is non-unique. In this case the cosmological horizon and the black hole horizon coincide which is analogous to the charged case with $\Lambda=0$. The inclusion of charge in the subsequent calculation will not change this situation. There exists the charged Nariai solution, see~\cite{BM} and references therein, which will saturate the inequality and thus uniqueness cannot be expected in this situation. In this section we investigate the sharpness issue when $Q\ne 0$ and $\Lambda>0$ and our main result in this section is similar to Proposition 1 in~\cite{AB}.

Let us consider a sequence of regular shell solutions which approach an infinitely thin shell. More precisely, by a regular solution $\Psi=(p,p_{\perp},\rho,q,a,b)$ of the Einstein equations we mean that $a$ and $b$ are $C^2$ except at finitely many points, that the quantities $p,p_{\perp},\rho$ and $q$ are $C^1$ except at finitely many points, $p$ has compact support and the Einstein equations are satisfied almost everywhere. Now let
\[
\Psi_k:=(p_k,(p_{\perp})_k,\rho_k,q_k,a_k,b_k)
\]
be a sequence of regular solutions such that
$p_k,(p_{\perp})_k,\rho_k$ and $q_k$ have support in $[R_0^k,R_1],$ where
\begin{equation}
  \lim_{k\to\infty}\frac{R_0^k}{R_1}=1.
  \label{hypothesis}
\end{equation}
Assume that
\begin{align}
  \label{supofp}
  \|r^2p_k\|_{\infty}<C,
\end{align}
and
\begin{align}
  \label{rhominuspperp}
  \int_{R_0^k}^{R_1}(\rho_k-2(p_{\perp})_k)r^2dr\to 0,\mbox{ as }k\to \infty.
\end{align}
Furthermore, assume that for some $\epsilon>0,$
\[
\frac{q_k^2}{r^2}+\Lambda r^2\leq 1-\epsilon, \mbox{ for }r\in [R_0^k,R_1].
\]
Finally, denote by $M_k$ and $Q_k$ the total gravitational mass and charge of the solution and assume that $M=\lim_{k\to\infty}M_k,$ and $Q=\lim_{k\to\infty}Q_k$ exist.
\begin{proposition}
\label{proposition}
Assume that $\{\Psi_k\}_{k=1}^{\infty}$ is a sequence of regular
solutions with the properties specified above.
Then
\begin{equation*}
  \frac{M}{R_1}=\frac29+\frac{Q^2}{3R_1^2}-\frac{\Lambda R_1^2}{3}+\frac29 \sqrt{1+\frac{3Q^2}{R_1^2} +3\Lambda R_1^2}-H(Q,\Lambda,R_1,M),
\end{equation*}
where $H\geq 0$ and $H=0$ if and only if $\Lambda R_1=0$ or $\frac{Q^2}{R_1^2}+\Lambda R_1^2=1.$
\end{proposition}
\textit{Remark 1: }
We note that sequences with the properties specified in the proposition has been
proved to exist for
the Einstein-Vlasov system in the case $Q=\Lambda=0,$ cf. \cite{An2}. It is interesting to note that the sequence of shells constructed in~\cite{An2}, which approach an infinitely thin shell, have support in $[R_0^j,R_0^j(1+(R_0^j)^q)],\, q>0,$ where $R_0^j\to 0$ as $j\to\infty.$ Hence, this sequence gives in the limit an infinitely thin shell at $r=0,$ which corresponds to the degenerate case $q^2/r^2+\Lambda r^2=0$ above.

\section{Proof of Theorem 1}
Inspired by the method of proof in~\cite{KS} we introduce new variables by
\begin{align*}
  x &= \frac{2m_g}{r} - \frac{q^2}{r^2} + \frac{\Lambda r^2}{3},\\
  y &= 8\pi r^2p,\\
  z_1 &= \frac{q^2}{r^2},\\
  z_2 &= \Lambda r^2.
\end{align*}
Note that conditions on $\rho$, $p$, $\Lambda$ and $q$ imply that our new variables belong to the set
\begin{equation}
  \label{U}
  {\cal{U}}:=\{(x,y,z_1,z_2): 0\leq x<1,\, y\geq 0,\, z_1\geq 0,\, z_2\geq 0,\, z_1+z_2\leq 1\}.
\end{equation}
The condition $x \neq 1$ excludes the charged Nariai class of solutions in the analysis. For an alternative derivation which includes the charged Nariai solutions see~\cite{BM}.

Einstein's equations can now be written in these new variables, we arrive at
\begin{align*}
  &8\pi r^2\rho = 2\dot{x} - x - z_1 - z_2,\\
  &8\pi r^2p = y,\\
  &8\pi r^2p_\perp = \frac{(x + y - z_1 - z_2)\dot{x}}{2(1 - x)} + \dot{y} - \dot{z_1} - z_1 + \frac{(x + y - z_1 - z_2)^2}{4(1 - x)},
\end{align*}
where $\dot{x}=\frac{dx}{d\beta}$ and $\beta = 2\log{r}$. Note that $\dot{z_2}=z_2$. Now condition (\ref{strong}) \(p + 2p_\perp \leq \rho \) becomes
\begin{equation*}
y + \frac{(x + y - z_1 - z_2)\dot{x}}{1 - x} + 2(\dot{y} - \dot{z_1} - z_1) + \frac{(x + y - z_1 - z_2)^2}{2(1 - x)}
 \leq 2\dot{x} + x - z_1 - z_2.
\end{equation*}
We can rearrange this and get
\begin{align*}
  \dot{x}(3x& + y - z_1 - z_2 - 2) + 2(\dot{y} - \dot{z_1} - z_2)(1 - x)\\
  \leq& -\frac{1}{2}(3x^2 + (y - z_1 - z_2)^2 - 2(x - y) - 2(z_2(4x - 3) + z_1)\\
  =: & -\frac{1}{2}u(x,y,z_1,z_2).
\end{align*}

We now define
\begin{equation*}
  w(x,y,z_1,z_2) = \frac{(4 - 3x + y - z_1 - z_2)^2}{1 - x}.
\end{equation*}
We will see that by determining the maximum of $w$ the claimed inequality will follow. Differentiating $w$ with respect to $\beta$ yields
\begin{align*}
  \dot{w}&=\frac{(4 - 3x + y - z_1 - z_2)}{(1 - x)^2}(\dot{x}(3x + y - z_1 - z_2 - 2) + 2(\dot{y} - \dot{z_1} - z_2)(1 - x))\\
  &\leq -\frac{1}{2}\frac{(4 - 3x + y - z_1 - z_2)}{(1 - x)^2}u(x,y,z_1,z_2).
\end{align*}
We note that $4 - 3x + y - z_1 - z_2\geq 0$ in view of (\ref{U}). Thus if $u(x,y,z_1,z_2)  \leq 0,$ then $w(x,y,z_1,z_2)$ is increasing. We can now determine $\sup_{{\cal{U}}}w,$ and thus require that $u \leq 0.$ This implies that
\begin{align}
  0 &\geq 3x^2 +(y - z_1 - z_2)^2 - 2(x - y) - 2(z_2(4x - 3) + z_1)\nonumber\\
  &= 3x(x - 1) + x - 8xz_2 + (y - z_1 - z_2)^2 + 2y - 2z_1 + 6z_2\nonumber\\
  &= (3x - 8z_2 + 1)(x - 1) + (y - z_1 - z_2 + 1)^2\nonumber\\
  &\implies (y - z_1 - z_2 + 1)^2 \leq (3x - 8z_2 + 1)(1 - x),\label{ineq1}
\end{align}
and this condition can be rearranged to give
\begin{align}
2(y-z_1 - z_2) &\leq (3x- 8z_2 + 1)(1-x) - 1 -(y-z_1 - z_2)^2 \nonumber\\
&\leq (3x- 8z_2 + 1)(1-x) - 1  \label{eqn:aa}
\end{align}
We get in view of (\ref{ineq1}) that
\begin{align*}
  w &= \frac{(4 - 3x + y - z_1 - z_2)^2}{1 - x}\\
  &= \frac{(1 + y - z_1 - z_2)^2}{1 - x} + 6(1 + y - z_1 - z_2) + 9(1 - x)\\
  &\leq 3x - 8z_2 + 1 + 6(1 + y - z_1 - z_2) + 9(1 - x)\\
  &= 16 - 6x + 6y - 6z_1 - 14z_2 = 16 -6x+ 6(y - z_1 - z_2) - 8z_2.
\end{align*}

Using equation~(\ref{eqn:aa}) leads to
\begin{align*}
  w &\leq 16 -6x + 3(3x - 8z_2 + 1)(1 - x) - 3 - 8z_2\\
  &= 16 - 9x^2 - 24z_2(1 - x) - 8z_2 \leq 16.
\end{align*}
Thus $\sup_{{\cal{U}}}w=16$. We immediately note from the definition of $w$ we that this value is attained for $x=y=z_1=z_2=0,$ i.e., $w(0,0,0,0,)=16.$ Since $w\leq 16$ in the domain ${\cal{U}}$ it follows in particular that $w\leq 16$ when $y=0,$ which implies
\begin{align*}
  &(3(1 - x) + 1 - z_1 - z_2)^2 \leq 16(1 - x)\\
  \implies &x \leq \frac{1}{9}(4 - 3z_1 - 3z_2) + \frac{4}{9}\sqrt{1 + 3z_1 + 3z_2}.
\end{align*}

We insert the expressions for $x$, $z_1$ and $z_2$ and rearrange slightly to finally get
\begin{equation*}
  \frac{m_g}{r} \leq \frac{2}{9} + \frac{q^2}{3r^2} - \frac{\Lambda r^2}{3} + \frac{2}{9}\sqrt{1 + \frac{3q^2}{r^2} + 3\Lambda r^2}\,,
\end{equation*}
which completes the proof of Theorem~\ref{thm1}.
\begin{flushright}
$\Box$
\end{flushright}

\section{Proof of Proposition \ref{proposition}}
Rewriting (\ref{match}) in terms of $m_g$ gives
\begin{equation}\label{brel}
e^{-b}=1-\frac{2m_g}{r}+\frac{q^2}{r^2}-\frac{\Lambda r^2}{3}.
\end{equation} 
We also reformulate (\ref{ee2}) and get
\begin{equation}\label{ar}
\frac{a_r}{2}=(4\pi r p+\frac{m_g}{r^2}-\frac{q^2}{r^3}-\frac{\Lambda r}{3})e^{b}.
\end{equation}
Below we drop the index $k$ but stress that the quantities $a,b,q,p,p_T,\rho$ and $R_0$
all depend on $k$ and in particular that $R_0^k/R_1\to 1$ as $k\to \infty.$
We define
\begin{align}
  \Gamma := (4\pi p r^3 + m_g -\frac{q^2}{r}- \frac{\Lambda r^3}{3})e^{(a+b)/2}
  \label{eqn:Gamma1}.
\end{align}
From the TOV equation together with the Einstein equations we then have
\begin{align}
  \Gamma' = (4\pi r^2(\rho + p + 2(p_{\perp}))+\frac{q^2}{r^2} - \Lambda r^2) e^{(a+b)/2}.
  \label{eqn:Gamma2}
\end{align}
Let us integrate Eq.~(\ref{eqn:Gamma2}) on the interval $[R_0,R_1].$ This leads to
\begin{eqnarray}
  \Gamma(R_1)-\Gamma(R_0) &=& \int_{R_0}^{R_1}\left[
  4\pi r^2 (\rho + p + 2p_{\perp})+\frac{q^2}{r^2} - \Lambda r^2
  \right] e^{(a+b)/2} dr\nonumber \\
  &=&\int_{R_0}^{R_1}8\pi r^2\rho\, e^{(a+b)/2}\,dr\nonumber\\
  & &+\int_{R_0}^{R_1}\left[
  4\pi r^2 (p + 2p_{\perp}-\rho)+\frac{q^2}{r^2} - \Lambda r^2
  \right] e^{(a+b)/2}\,dr\nonumber\\
  &=:&S+T.
\end{eqnarray}
In view of the assumptions on the sequence, and that $e^{(a+b)/2}\leq 1,$ we find that $T=O(|R_1-R_0|),$ and in particular $T\to 0$ as $k\to\infty.$
For the term $S$ we have by the mean value theorem for integration
\begin{eqnarray}
S&=& 2e^{a(\xi)/2}\xi \int_{R_0}^{R_1}4\pi r\rho\, e^{b/2}\,dr\nonumber \\
  &=&-2e^{a(\xi)/2}\xi \int_{R_0}^{R_1}\frac{d}{dr}e^{-b(r)/2}\,dr\nonumber\\
  & &+2e^{a(\xi)/2}\xi\int_{R_0}^{R_1}\left[
  \frac{m_g(r)}{r^2}-\frac{q^2}{r^3}-\frac{\Lambda r}{3} \right] e^{b/2}\,dr=:
S_1+S_2,\nonumber
\end{eqnarray}
where $\xi\in]R_0,R_1[.$
From the assumption that $q^2/r^2+\Lambda r^2\leq 1-\epsilon,$ it follows that
$e^{b/2}\leq C(\epsilon).$ Indeed, by (\ref{brel})
\[
e^{b}=\frac{1}{1-2m_g/r+q^2/r^2-\Lambda r^2/3},
\]
and in view of our main inequality we have
\begin{eqnarray}\label{sigma}
\frac{2m_g}{r}-\frac{q^2}{r^2}+\frac{\Lambda r^2}{3}&\leq& 4/9-(q^2/r^2+\Lambda r^2)/3+\frac49
\sqrt{1+3(q^2/r^2+\Lambda r^2)}\nonumber\\
&=&4/9-\sigma/3+\frac49\sqrt{1+3\sigma},
\end{eqnarray}
with $\sigma=q^2/r^2+\Lambda r^2.$ It is easy to see that the right hand side is
bounded by $1$ and strictly less than one if $\sigma<1,$ which shows that
$e^{b/2}\leq C(\epsilon)$ as claimed.
Now, since
\[|m_g/r-q^2/r^2-\Lambda r^2/3|\leq C,
\]
we get
\[
S_2\leq C\log{(R_1/R_0)}=O(|R_1-R_0|).
\]
In view of (\ref{ar}) and the assumption that $\|r^2 p\|_{\infty}\leq C,$ the same estimates show that
\[
e^{a(\xi)/2}\to \sqrt{1-\frac{2M}{R_1}+\frac{Q^2}{R_1^2}-\frac{\Lambda R^2_1}{3}},
\mbox{ as }\xi\to R_1.
\]
By introducing the notation
\[
\Omega:=\sqrt{1-\frac{2M}{R_1}+\frac{Q^2}{R_1^2}-\frac{\Lambda R^2_1}{3}},
\]
we now get by evaluating $S_1,$
\begin{eqnarray}
\Gamma(R_1)&=&\Gamma(R_0)+
2R_1\Omega\left(\sqrt{1-\frac{\Lambda R^2_1}{3}}-\Omega\right)+O(|R_1-R_0|)\nonumber\\
&=&-\frac{\Omega\Lambda R_1^3}{3\sqrt{1-\frac{\Lambda R_1^2}{3}}}+2R_1\Omega\left(\sqrt{1-\frac{\Lambda R^2_1}{3}}-\Omega\right)+O(|R_1-R_0|)\nonumber\\
&=&2R_1\Omega\left(1-\Omega\right)-\frac{\Omega\Lambda R_1^3}{3}\left(\frac{2}{1+\sqrt{1-\frac{\Lambda R_1^3}{3}}}+\frac{1}{\sqrt{1-\frac{\Lambda R_1^3}{3}}}\right)\nonumber\\
& &+O(|R_1-R_0|).
\end{eqnarray}
In the limit we therefore obtain, using that $(a+b)(R_1)=0,$
\begin{equation}\label{limith}
\frac{\Gamma(R_1)}{R_1}=\frac{M}{R_1}-\frac{Q^2}{R_1^2}-\frac{\Lambda R_1^2}{3}=2\Omega\left(1-\Omega\right)-h,
\end{equation}
where
\[
h:=\frac{\Omega\Lambda R_1^2}{3}\left(\frac{2}{1+\sqrt{1-\frac{\Lambda R_1^3}{3}}}+\frac{1}{\sqrt{1-\frac{\Lambda R_1^3}{3}}}\right).
\]
A straightforward computation shows that (\ref{limith}) is equivalent to the equation
\begin{eqnarray}
&9&\left(\frac{M}{R_1}-\frac{2}{9}-\frac{Q^2}{3R_1^2}+\frac{\Lambda R_1^2}{3}-\frac{2}{9}\sqrt{1+\frac{3Q^2}{R_1^2}+3\Lambda R_1^2}\,\right)\nonumber\\
& &\times\left(\frac{M}{R_1}-\frac{2}{9}-\frac{Q^2}{3R_1^2}+\frac{\Lambda R_1^2}{3}+\frac{2}{9}\sqrt{1+\frac{3Q^2}{R_1^2}+3\Lambda R_1^2}\,\right)=h^2-4h\Omega.\nonumber
\end{eqnarray}
We note that
\[
M=\int_{R_0}^{R_1}4\pi\eta^2\rho\, d\eta+\frac{Q^2}{2R_1}+\frac{f(R_1)}{2},
\]
which implies that the second factor is strictly positive when $\rho\ne 0.$
Thus we can write
\begin{eqnarray}
\frac{M}{R_1}&=&\frac{2}{9}+\frac{Q^2}{R_1^2}-\frac{\Lambda R_1^2}{3}+\frac{2}{9}\sqrt{1+\frac{3Q^2}{R_1^2}+3\Lambda R_1^2}\nonumber\\
& &-\frac{h(4\Omega-h)}{9\left(\frac{M}{R_1}-\frac{2}{9}-\frac{Q^2}{R_1^2}+\frac{\Lambda R_1^2}{3}+\frac{2}{9}\sqrt{1+\frac{3Q^2}{R_1^2}+3\Lambda R_1^2}\,\right)}\nonumber\\
&=:&\frac{2}{9}+\frac{Q^2}{R_1^2}-\frac{\Lambda R_1^2}{3}+\frac{2}{9}\sqrt{1+\frac{3Q^2}{R_1^2}+3\Lambda R_1^2}-H.\label{equalityH}
\end{eqnarray}
Now $H\geq 0$ since $\Lambda R_1^2\leq 1,$ and $H=0$ if and only if
$\Lambda R_1=0$ or $\Omega=0.$ If $\Omega=0$ then
\begin{equation}\label{Hequal0}
1-\frac{2M}{R_1}+\frac{Q^2}{R_1^2}-\frac{\Lambda R^2_1}{3}=0.
\end{equation}
In view of (\ref{sigma}) this implies that
\[
\frac{Q^2}{R_1^2}+\Lambda R^2_1=1.
\]
\begin{flushright}
$\Box$
\end{flushright}

\section{Summary}
We have investigated spherically symmetric charged objects in the case of a
positive cosmological constant and we have derived a bound on the ratio of
the gravitational mass $m_g$ and the area radius $r$ given by
\begin{equation}\label{mainineqs}
\frac{m_g}{r} \leq \frac{2}{9} + \frac{q^2}{3r^2} - \frac{\Lambda r^2}{3} + \frac{2}{9}\sqrt{1 + \frac{3q^2}{r^2} + 3\Lambda r^2},
\end{equation}
under the assumption that
\begin{equation}\label{condition}
0\leq \frac{q^2(r)}{r^2}+\Lambda\,r^2\leq 1.
\end{equation}
The bound (\ref{mainineqs}) is very general in the context of spherically
symmetric solutions. However, in the case when $\Lambda=0$,
the bound is sharp and sharpness is obtained by infinitely
thin shell solutions~\cite{An5} but
if $\Lambda>0$ we have shown that
infinitely thin shell solutions \textit{do not} in general saturate 
the inequality (\ref{mainineqs}). Moreover, if (\ref{condition}) is not
satisfied the problem is open and this includes in particular the case of a negative cosmological
constant.

\end{document}